# Optically-stimulated desorption of "hot" excimers from pre-irradiated Ar solids

G.B. Gumenchuk[1,2], I.V. Khyzhniy[1], A.N. Ponomaryov[2],

M.A. Bludov[1], S.A. Uyutnov[1], A.G. Belov[1], E.V. Savchenko[1], V.E. Bondybey[2]

[1] *Institute for Low Temperature Physics & Engineering NASU, Lenin Ave 47, 61103 Kharkov, Ukraine*

[2] *Lehrstuhl für Physikalische Chemie II TU München, Lichtenbergstraße 4, 85747 Garching, Germany*

e-mail: savchenko@ilt.kharkov.ua

**Abstract**

Electronically-induced desorption from solid Ar pre-irradiated by a low-energy electron beam was investigated by activation spectroscopy methods - photon-stimulated exoelectron emission and photon-stimulated luminescence in combination with spectrally-resolved measurements in the VUV range of the spectrum. Desorption of vibrationally excited argon molecules $Ar_2^{*(v)}$ from the surface of pre-irradiated solid Ar was observed for the first time. It was shown that desorption of "hot" $Ar_2^{*(v)}$ molecules is caused by recombination of self-trapped holes with electrons released from traps by visible range photons. The possibility of optical stimulation of the phenomenon is evidenced.

PACS: 79.20.La Photon- and electron-stimulated desorption

*Keywords:* photon stimulated desorption, electron stimulated desorption, exoelectron emission.

# 1. Introduction

Desorption from rare gas crystals was studied under excitation by ions [1,2], electrons [3,4] and photons [5,6]. Desorbing particles are ground state atoms, excited atoms, excited molecules and ionic species. Desorption stimulated by electronic excitation of the crystal was first observed in experiments on the diffraction of slow electrons [7]. The basis of this phenomenon is the conversion of electronic excitation energy into kinetic energy of atoms caused by the localization of excitations near the surface of the crystal.

When a rare gas solid (RGS) is irradiated by ionizing radiation with excitation energy higher than the forbidden energy gap, electron-hole pairs are produced. Electrons behave like free particles in RGS (except He), while holes are self-trapped in a short period of time - $10^{-12}$ sec [8].

Let us consider the desorption from RGS on the example of solid Ar. Excimer molecules in Ar lattice are produced due to self-trapping of excitons or due to recombination of self-trapped holes with electrons followed by the formation of self-trapped excitons. The transition of the excited molecule to the repulsive part of the ground state term is accompanied by emission of the broad M-band with a maximum at 9.7 eV.

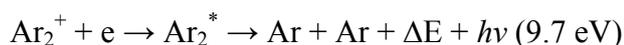
$$Ar_2^+ + e \to Ar_2^* \to Ar + Ar + \Delta E + h\nu \ (9.7\ eV)$$

This recombination reaction of self-trapped holes with electrons stimulates the desorption of Ar atoms from solid Ar to the vacuum. The energy $\Delta E$ released in this process is approximately 1 eV (~0.5 eV per atom). This value is enough for an atom to overcome the energy barrier and escape from the sample to the vacuum in case recombination occurs on the surface of the crystal – excimer dissociation mechanism of desorption [9,10]

Excited Ar atoms desorb from the surface of solid Ar because of repulsive interaction between an excited electron and surrounding atoms of a regular lattice caused by negative electron affinity of crystalline Ar [8] – the so-called "cavity-ejection" mechanism [9]. The partial contribution of excited atoms to the desorption was distinguished in [4,6] and was shown to be stimulated by direct generation of excitons in the crystal. In [11] a mechanism of desorption of excited atoms from solid Ar under low-energy electron beam excitation was discussed. Experiments were performed using luminescence VUV spectroscopy. It was shown that nonthermalised excitons play the main role in the transport of excitation energy to the surface.

The desorption of "hot" molecules from rare gas solids was observed before only during excitation of the crystals. Electronically induced desorption of vibrationally hot molecules from surfaces of solid Ar [12] and Ne [13,14] was identified by observation of luminescence stemming from a plume of sputtered particles under excitation. Excimer desorption from solid Ar under selective excitation in the range of 10-35 eV was studied in [15]. Excitation spectra of the W-band demonstrated that primary creation of excitonic states is necessary for ejection of Ar excimers and electron-hole recombination stimulates this process as well. The formation of excimers is caused by the self-trapping of excitons in the lattice due to exciton-phonon interaction. Desorption of hot molecules $Ar_2^{*(v)}$ from solid Ar can occur via different mechanisms. First one is the "cavity-ejection" mechanism operating in RGS with negative electron affinity [8,9]. In solid Ar it causes repulsion between excited particles and surrounding atoms of regular lattice. Therefore a cavity around $Ar_2^*$ is formed. Thus excimers formed on the surface of the sample are pushed out to the vacuum. Another mechanism is concerned with lattice rearrangements in the vicinity of excimer. Appearance of the dimer in the interstitial position is followed by its shift along the <110> direction in the lattice. The electronic excitation energy is transferred to a specific motion of the dimer. This motion can result in desorption of excimer when the described process occurs on the surface. The energy needed for such dimer motion can also be released in the lattice in the course of the vibronic relaxation. A theoretical investigation [16] showed that in a system with strong local vibration the energy release proceeds in a jump-like multiphonon process.

In case of pre-irradiated rare gas solids initial states of relaxation cascades are self-trapped holes in molecular-type configuration $Rg_2^+$ and trapped electrons. It was found recently in a molecular dynamics study of energy transfer in solid Ar that the energy can be transferred over long distances from the excitation site [17]. Desorption of atoms in the ground state induced by charge recombination in the bulk of the crystal can take place via the so-called crowdion mechanism, suggested in [18] to explain anomalous low-temperature desorption of Ar atoms from solid Ar pre-irradiated by an electron beam. Crowdions are non-linear waves of atomic displacements, which propagate through the crystal to comparatively long distances (~100 of lattice constants). These quasi-particles can exist in solid Ar, as was shown in [19], and can take part in the desorption of atoms from the sample transmitting energy to the sample surface. The authors calculated the energy needed for the creation of crowdion to be 0,3 eV, which is lower than that released due to the radiative decay of the excited dimer $Ar_2^*$. This process is the main stimulating factor for desorption of Ar atoms at temperatures much lower than the characteristic sublimation temperature for solid Ar (30 K).

In this article we present the first observation of a nontrivial post-irradiation phenomenon - desorption of vibrationally "hot" excimers $Ar_2^*$ from solid Ar pre-irradiated by an electron beam.

## 2. Experiment

Before the experiment the gas inlet system was pumped and degassed by heating under pumping. The samples were condensed from the gas phase under isobaric conditions (P = $10^{-7}$ mbar) on a metal substrate cooled by a closed-cycle 2-stage Leybold RGD 580 cryostat to the temperature 9K. High purity Ar (99,999%) was used. The base pressure in the vacuum chamber was $5\times10^{-8}$ mbar. A typical sample thickness was 100 μm. The samples were irradiated with an electron beam to generate electron-hole pairs. We used electrons of 500 eV energy and the current density of 30 μAcm$^{-2}$. The dose of irradiation was increased with exposure time.

To register cathodoluminescence spectra of solid Ar a VUV monochromator was used. We performed spectrally resolved measurements in the VUV range to detect emission from the bulk of the crystal and from the desorbing particles. The measurements of luminescence intensity were performed in a photon-counting mode. After the irradiation was finished the yields of afterglow and afteremission of electrons were measured. The emission of electrons from pre-irradiated samples was detected with an Au-coated Faraday plate kept at a small positive potential +9 V. It was positioned in front of the sample at the distance of 10mm. The current from the Faraday plate was amplified by a FEMTO DLPCA 100 current amplifier.

For experiments on photon stimulated exoelectron emission we used a Coherent 899-05 dye laser pumped with an Ar-ion laser and tuned to 633 nm wavelength. The power of the laser beam was 4 mW. The sample heating under laser light did not exceed 0.5 K. In photo-induced recombination luminescence experiments a vacuum monochromator was tuned to the wavelength corresponding to the W-band of solid Ar. Then the pre-irradiated sample was exposed to a He-Ne laser beam of 4 mW power and 632,8 nm wavelength.

### 3. Results and discussion

During irradiation of the samples of solid Ar by an electron beam the spectra of cathodoluminescence were recorded. The luminescence spectrum of nominally pure solid Ar under irradiation with an electron beam consists of features that originate from the bulk and surface-related features. The most intense luminescence is caused by the radiative decay of molecular-type self-trapped excitons Ar$_2^*$ in the bulk of the sample. This broad so-called M-band is formed due to transitions of excimers from $^{1,3}\Sigma_u^+$ states to the ground state $^1\Sigma_g^+$ and its intensity is much higher than that of other features in the spectrum. In figure 1 the cathodoluminescence spectrum of solid Ar in the range of interest (10.7-11.8 eV) under irradiation by low-energy electrons is presented. The well-known M-band is not shown here. Band c is ascribed to atomic-type excitons self-trapped on defect sites in the bulk of the sample and is analog to the $^3P_2 \rightarrow {}^1S_0$ transition in the free Ar atom. The sharp line b is emitted by the excited Ar$^*$ atoms desorbed from the surface of the sample to the vacuum in $^3P_1$ state. It is due to a transition to the ground state $^3P_1 \rightarrow {}^1S_0$. The line corresponding to transition $^1P_1 \rightarrow {}^1S_0$ is too weak in this case. The intensity distribution between transitions from the $^3P_1$ and $^1P_1$ states is inversed in comparison with that in the gas phase. The population of the triplet state is preferable because of interaction with phonons in the lattice. The W-band stems from the desorbing excimer molecules $Ar_2^{*(v)}$ as a result of transitions from vibrationally excited states $^3\Sigma_u^{+(v)}$ to the ground $^1\Sigma_g^+$ state. Note that vibrational relaxation on the surface is much slower than in the bulk. Therefore there is a high probability for molecules to desorb from the surface in a "hot" state.

After the irradiation of a sample was completed we started to investigate the relaxation processes in the crystal. If solid Ar contains a minor amount of nitrogen a long-lifetime afterglow is observed after switching off the irradiating electron beam. This effect is caused by the presence of guest nitrogen atoms (from residual gases in the vacuum chamber) in metastable states after excitation with electrons. The well-known transition of the N atom $^2D \rightarrow {}^4S$ is followed by the emission of light with a 521 nm wavelength. We found a similar effect in the yield of exoelectron current – afteremission of electrons from the sample on completion of irradiation [20]. It was shown that afterglow is responsible for afteremission of exoelectrons. The current decays exponentially in time. The decay curves can be described by a second-order exponential function with characteristic decay times $\tau_1 \sim 30$ sec, $\tau_2 \sim 170$ sec [21]. We observed a similar dependence of the M-band [22] – afterglow in the VUV range of spectrum, stemming from the bulk self-trapped excitons. The estimated decay time for the afterglow of the M-band is about 19 seconds, which is close to the characteristic lifetime of the radiative transition of N atom from the metastable state $^2D$ to the $^4S$ state in Ar matrix [23].

During the afteremission process we switched on the laser beam directed to the surface of the sample to register the release of electrons from the traps by visible light. Note that holes are self-trapped in RGS, electrons can be trapped either by guest atoms or by such lattice defects as vacancies or pores (in view of the negative electron affinity of solid Ar). Taking into account that solid Ar possess quite a wide conduction band (several eV) one can expect to release electrons from the traps (both deep and wide ones) using visible light. In figure 2 (a) the yield of laser-induced exoelectron emission from pre-irradiated solid Ar recorded during afteremission is presented. The release of electrons from the traps and the following escape from the sample under the laser light is obvious. Some fraction of free electrons reaches the surface of the sample and goes to the vacuum due to the absence of an energy barrier because of negative electron affinity. Other electrons can either recombine with self-trapped holes or positively charged guest atoms or can be retrapped.

In this study we measured the yield of afterglow on the wavelength of the W-band emitted by desorbed $Ar_2^{*(v)}$ (fig. 2b). It was found that the intensity of this emission also decayed exponentially with a characteristic decay time $\tau \sim 2.5$ seconds. The reason why the value of $\tau$ is much smaller for the W-band in comparison with that for the M-band lies in the origin of these bands. The M-band is emitted by the centers in the bulk of the crystal while the W-band is caused by excimers formed on the surface and in its vicinity. The rate of surface centers depletion is higher than that for the bulk ones.

It is most likely that the radiative recombination of self-trapped holes with electrons can be considered as a source of the hot molecule desorption from solid Ar after preliminary irradiation of the crystal with an electron beam. Then the recombination of electrons with self-trapped holes should increase the number of hot $Ar_2^{*(v)}$ molecules, desorbed from the sample. We detected these particles spectroscopically by measuring the intensity of W-band. Figure 2b demonstrates the effect of photo-induced luminescence of solid Ar at the wavelength corresponding to the W-band (photon energy 11.3 eV). The laser light (wavelength 632.8 nm) was focused on the surface of the sample. The yield of luminescence was recorded immediately after the irradiating electron beam was switched off. The intensity of the W-band started to decrease exponentially. Then the laser was switched on after approximately 5 seconds. We observed a considerable increase in W-band intensity and a following exponential decay, caused by the depopulation of electron traps by laser light. This fact proves the suggestion that electron-hole recombination is responsible for the desorption of hot molecules from pre-irradiated solid Ar. The increase in W-band intensity (number of desorbed excimers) induced by laser light depends on the moment when the laser was switched on. The longer the delay between switching off the electron beam and switching on the laser, the lower the intensity of the photon-stimulated W-band emission.

The results obtained give evidence of a possibility of optical stimulation of post-irradiation desorption of "hot" $Ar_2^{*(v)}$ dimers from the surface of solid Ar containing self-trapped holes, preliminarily produced by irradiation of the crystal with some kind of ionizing radiation.

## Summary

In this study we observed for the first time the desorption of excited molecules $Ar_2^{*(v)}$ from pre-irradiated solid Ar. It was shown that desorption of vibrationally "hot" eximers from the solid Ar after preliminary excitation above the forbidden energy gap is stimulated by recombination of electrons with self-trapped holes. It is demonstrated that the process can be triggered optically by releasing the electrons from the traps.

**Acknowledgements**

The authors thank Professors P. Feulner and G. Zimmerer for valuable discussions. Financial support from Deutsche Forschungsgemeinschaft is gratefully acknowledged.

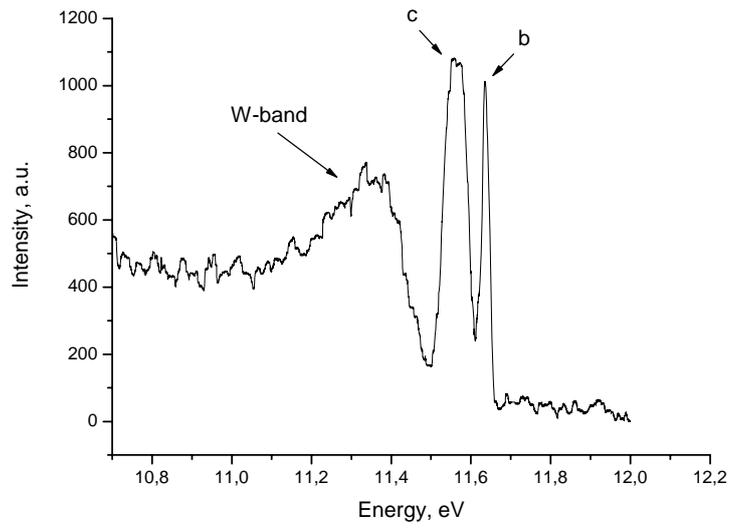

Fig. 1

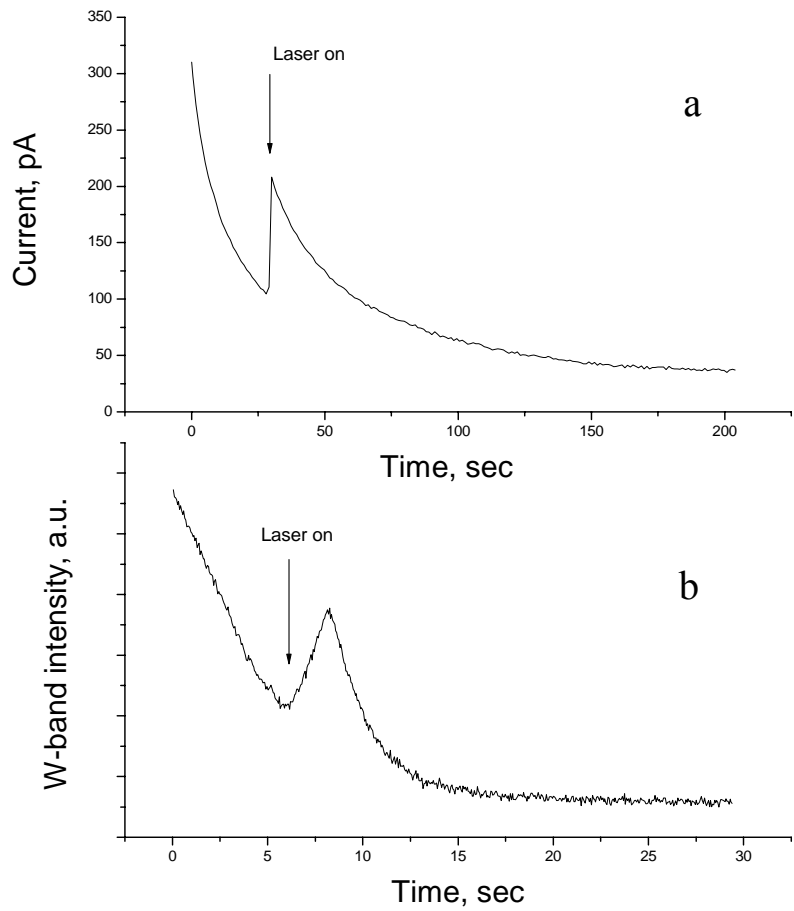

Fig. 2

**Figure captions**

Fig. 1. Cathodoluminescence spectrum of solid Ar in the range 10.7-11.8 eV.

Fig. 2. a - Laser-induced exoelectron emission current from pre-irradiated solid Ar recorded during afteremission.

b - Laser-induced luminescence of solid Ar on the wavelength corresponding to emission of desorbed hot molecules $Ar_2^*$ (W-band)

Sample temperature T = 9K.